\documentclass[preprint2]{aastex}

\slugcomment{Prepared for ApJ}
\shorttitle{Transit timing variation for coorbiting planets}
\shortauthors{}

\begin{document}

\title{Transit timing variations for planets coorbiting \\ in the horseshoe regime}

\author{David Vokrouhlick\'y}
\affil{Institute of Astronomy, Charles University,
       V Hole\v{s}ovi\v{c}k\'ach 2, CZ--18000 Prague 8, \\
       Czech Republic, E-mail: vokrouhl@cesnet.cz}
\author{David Nesvorn\'y}
\affil{Department of Space Studies, Southwest Research Institute,
      1050 Walnut St., Ste 300, \\ Boulder, CO 80302,
      E-mail: davidn@boulder.swri.edu}

\begin{abstract}
 While not detected yet, pairs of exoplanets in the 1:1 mean motion resonance
 probably exist. Low eccentricity, near-planar orbits, which in the
 comoving frame follow the horseshoe trajectories, are one of the possible
 stable configurations. Here we study transit timing variations
 produced by mutual gravitational interaction of planets in this orbital
 architecture, with the goal to develop methods that can be used to recognize
 this case in observational data. In particular, we use a semi-analytic model
 to derive parametric constraints that should facilitate data analysis. We
 show that characteristic traits of the transit timing variations can directly
 constrain the (i) ratio of planetary masses, and (ii) their total mass (divided
 by that of the central star) as a function of the minimum angular separation
 as seen from the star. In an ideal case, when transits of both planets are
 observed and well characterized, the minimum angular separation can also be
 inferred from the data. As a result, parameters derived from the observed transit timing
 series alone can directly provide both planetary masses scaled to the central
 star mass.
\end{abstract}


\section{Introduction}
Stable configurations of planets in the 1:1 mean motion resonance 
(MMR) comprise three different cases: (i) tadpole orbits, which are
similar to the motion of Trojan asteroids near Jupiter's L4 and
L5 stationary points, (ii) horseshoe orbits, which are similar
to the motion of Saturn's satellites Janus and Epimetheus, and (iii)
binary-planet orbits, in which case the two planets revolve about
a common center-of-mass moving about the star on a Keplerian
orbit. Numerous studies explored these configurations with
different aims and goals. Some mapped stability zones in 
orbital and parametric spaces. Other studies dealt with
formation and/or capture of planets in the 1:1 MMR
and their survivability during planetary migration.
Still other works explored observational traits such as the radial
velocity (RV) signal in stellar spectrum or transit timing variations
(TTVs) in the case that one or both planets transit the stellar disk.
While we still do not know details of dominant formation and
evolutionary processes of planetary systems, as well their
variety, a general consensus is that planets in 1:1 MMR 
should exist. Here we briefly recall several important studies
directly related to our work.

Laughlin \& Chambers (2002), while studying methods that
would reveal a pair of planets in 1:1 MMR from the RV analysis
(see also Giuppone et~al. 2012), pointed out
two possible formation scenarios: (i) planet-planet scattering that
would launch one of the planets into a coorbital zone of another
planet (including possibly one of the high-eccentricity stable 
orbital configurations%
\footnote{Note that there is a surprising variety of 1:1 MMR 
 planetary configurations, many of which have large eccentricities
 or inclination (e.g., Giuppone et~al. 2009, 2012; Hadjidemetriou
 et.~al. 2009, Schwarz et~al. 2009, Hadjidemetriou \& Voyatzis 2011,
 Haghighipour et~al. 2013, Funk et~al. 2013). In this paper, we
 do not consider these cases.}),
and (ii) in-situ formation of a smaller planet near the L4 or
L5 points of a Jupiter-class planet. These authors also noted
that the 1:1 MMR would persist during subsequent migration, since
the balance between angular momentum and energy loses prevents an 
eccentricity increase. This behavior stands in contrast with planets 
captured in other (higher-order) resonances. Moreover, 
if significant gas drag is present, the libration amplitude may be 
damped, thus stabilizing the coorbital configuration.

The suggested scenario of in-situ formation by Laughlin 
\& Chambers (2002) has been modeled by several groups. Beaug\'e
et~al. (2007) started with a population of sub-lunar mass
planetesimals already assumed to be present in the tadpole region 
of a giant planet, and studied conditions of coorbiting planet 
growth. They took into account mutual gravitation interaction of the 
planetesimals as well as several gas-density models. With this 
set-up, Beaug\'e et~al. (2007)
noted that only $\simeq 0.6$ Earth mass planets grow in
their simulations. Beaug\'e et~al. also conducted simulations of
planet growth during the migration phase and found essentially the
same results. Notably, the Trojan planet orbit
has not been destabilized and safely survived migration with
a low final eccentricity. 

A more detailed study has been presented
by Lyra et~al. (2009). Using a sophisticated model of gas and solid 
dynamics in a self-gravitating thin disk, these authors modeled
planet formation starting with centimeter-size pebbles. They
showed that pressure maxima associated with macroscopic vortexes
may collect enough particles to generate instability followed by
gravitational collapse. Up to $5-20$ Earth mass planets may form 
this way in the tadpole 
region of a Jupiter-mass primary, depending on size distribution 
of the pebble population. 

Another pathway toward the formation of low-eccentricity planets in 
the 1:1 MMR
has been studied by Cresswell \& Nelson (2006). These authors analysed the
orbital evolution of a compact system of numerous Earth- to super-Earth-mass 
planets during the dynamical instability phase. If the planetary orbits
were initially several Hill radii apart in their simulations, the coorbital
configuration emerged as a fairly typical case for some
of the surviving planets. A similar set-up, though with different
assumptions about the gas-disk density profile, has also been studied
by Giuppone et~al. (2012), who showed that even similar-mass planets
may form in the 1:1 MMR configuration. Additionally, these authors
also simulated the coorbital-planet formation and stability during the 
gas-driven migration. Common to these works was that the 
1:1 MMR configuration formed in a sufficiently low-eccentricity state, 
assisted by efficient gas friction, prior or during the migration
stage. If, on the other hand, most of close-in planets formed as
a result of tidal evolution from a high-eccentricity state, acquired
during the planet-planet scattering (e.g., Beaug\'e \& Nesvorn\'y
2012), and the gas drag was of no help to keep the eccentricities low
at any stage of evolution, the fraction of surviving 1:1 MMR
configurations may be very small.

Orbital evolution and survival of planets in the 1:1 MMR during  
migration has also been studied in some detail. For instance, Cresswell
\& Nelson (2009) considered dynamics of coorbiting planets during
and after the gas-disk dispersal and generally found the
system to be stable. In some cases, the late migration stage with
low-gas friction, or after nebula dispersal, has resulted 
in an increase of the libration amplitude of the tadpole regime and 
transition into a horseshoe regime, or even destabilization
(see also analysis in Fleming \& Hamilton 2000). Rodr{\'\i}guez et~al.
(2013) included also tidal interaction with the star and found
that equal-mass planets may suffer destabilization during their
inward migration. Unequal-mass configurations on the other hand that
naturally form in the in-situ scenario, may thus be more common.

As far as the detection methods are concerned, the easiest idea
would be to seek photometric dips about 1/6 for the hot-Jupiter
orbital period away from its transit (as expected for a planet located
in the Lagrangian stationary points L4 or L5). However, this
approach did not yield so
far a positive result (e.g., Rowe et~al. 2006, Moldovan et~al.
2010). For that reason researchers sought other detection 
strategies. For instance, 
Ford \& Gaudi (2006) found that a Trojan companion to hot Jupiter
might be revealed by detecting an offset between the mid-time
of its transit and the zero point of the radial velocity of the star
(assuming that barycenter motion is subtracted). This effect would be 
detectable with available technology for planet companions with at
least several Earth masses. While interesting, this method requires
a combination of high-quality TTV and RV observations. So far,
only upper limits of putative Trojan companions were obtained with
this method.

A method based uniquely on analysis of the TTVs of hot Jupiter,
if accompanied by a Trojan planet, was discussed by Ford \& Holman
(2007). While finding the TTV amplitude large enough for even
low-mass Trojan companions, Ford \& Holman (2007) also pointed out
difficulties in interpretation of the data. For instance a Trojan 
planet on a small-amplitude tadpole orbit would produce nearly
sinusoidal TTVs in the orbit of giant planet. Such signal may be
produced by a distant moon and/or resonant perturbations due to
additional planets in the system. It would take further tests and
considerations to prove the signal is indeed due to a Trojan
companion.

Haghighipour et~al. (2013) presented so far the most detailed study
of TTVs produced by a Trojan companion of transiting hot Jupiter. 
Their main goal was to demonstrate that the expected TTV amplitudes were
within the detectable range of Kepler (or even ground-based)
observations. With that goal, they first numerically determined the 
stable region in the orbital phase space. Next, they modeled the TTVs
in the hot Jupiter orbit, giving several examples of how the amplitude
depends on key parameters of interest (mass of the Trojan
companion and eccentricity of its orbit, orbital period of
hot Jupiter, etc.). While confirming a confidence of detectability 
of the produced TTVs, this work did not give any specific hints about
inversion problem from TTVs to the system's parameters neither it
discussed uniqueness of the TTV-based determination of Trojan-planet
properties.

In this work, we approach the problem with 
different tools. Namely, we develop a semi-analytic 
perturbative method suitable for low-eccentricity orbits
in the 1:1 MMR. While our method can be applied to the tadpole regime,
or even the binary planet configuration, we discuss the case of
a coorbital planets on horseshoe orbits. This is because in this
case the TTV series have a characteristic shape, which would
allow us to most easily identify the orbital configuration (see also
Ford \& Holman 2007). While the final TTV inversion problem needs
to be performed numerically, multi-dimensionality of the parameter
space is often a problem.
Our formulation allows us to set approximate constraints on several
key parameters such as the planetary masses and amplitude of the 
horseshoe orbit. This information can be used to narrow the volume of
parameter space that needs to be searched. Analytic understanding of
TTVs is also useful to make sure that a numerical solution is physically
meaningful.

\section{Model}
Following Robutel \& Pousse (2013), we use the Poincar\'e relative 
variables $({\bf r}_0,{\bf r}_1,{\bf r}_2; {\bf p}_0,{\bf p}_1,
{\bf p}_2)$ to describe motion of the star with mass $m_0$ and 
two planets with masses $m_1$ and $m_2$. It is understood
that $(m_1,m_2)\ll m_0$.
The stellar coordinate ${\bf r}_0$ is given by its position
with respect to the barycenter of the whole system, and the
conjugated momentum ${\bf p}_0$ is the total (conserved) linear
momentum of the system. Conveniently, ${\bf p}_0$ is set to be 
zero in the barycentric inertial system. The coordinates
$({\bf r}_1,{\bf r}_2)$ of planets are given by their relative position with 
respect to the star, and the conjugated momenta $({\bf p}_1,{\bf p}_2)$
are equal to corresponding linear momenta in the barycentric
frame. The advantage of the Poincar\'e variables stems from
their canonicity (e.g., Laskar \& Robutel 1995, Go\'zdziewski
et~al. 2008). Their slight caveat is that the coordinates and momenta
are given in different reference systems, which can produce
non-intuitive effects (see, e.g., Robutel \& Pousse 2013). These
are, however, of no concern in our work.

Heading toward the perturbation description, the 
total Hamiltonian ${\cal H}$ of the system is divided into the 
unperturbed Keplerian part
\begin{equation}
 {\cal H}_{\rm K} = \sum_{i=1}^2 \left(\frac{p_i^2}{2\,
  \mu_i}-G\,\frac{\mu_i M_i}{r_i}\right)\; ,
 \label{ham1}
\end{equation}
and the perturbation
\begin{equation}
 {\cal H}_{\rm per} = \frac{{\bf p}_1\cdot{\bf p}_2}{m_0}-
  G\,\frac{m_1 m_2}{\left|{\bf r}_1-{\bf r}_2\right|}\; .
  \label{ham2}
\end{equation}
Here we denoted $M_i=m_0+m_i$ and the reduced masses $\mu_i=m_0 m_i/
M_i$ for the planets $i=1,2$. The gravitational constant is denoted by
$G$.

For sake of simplicity, we restrict the analysis to the planar 
configuration. The reference plane of the coordinate system is then chosen
to coincide with the orbital plane of the two planets around the star.
As a result, the planetary orbits are described by only
four orbital elements: semimajor axis $a$, eccentricity $e$, longitude
of pericenter $\varpi$ and mean longitude in orbit $\lambda$. To preserve
canonicity of the orbital parameters, and to deal with orbits of 
small eccentricity, we adopt Poincar\'e rectangular variables 
$(\lambda,\Lambda;x,-\imath \bar{x})$, instead of the simple Keplerian 
set, to describe orbits of both planets ($\imath=\sqrt{-1}$ and
over-bar meaning complex conjugate operation). Here the 
momentum conjugated to the longitude of orbit $\lambda$ is the Delaunay 
variable $\Lambda = \mu\,\sqrt{GM a}$. The complex coordinate $x=\sqrt{\Lambda}\,
\sqrt{1-\sqrt{1-e^2}}\,\exp(\imath \varpi)$ has its counterpart in
the momentum $-\imath \bar{x}$, both fully describing eccentricity
and pericenter longitude. In a very small eccentricity regime we may also
use a non-canonical, but simpler, variable $z=e\,\exp(\imath \varpi)
= \sqrt{2/\Lambda}\, x + O(x^3)$.

Since the difference in mean longitudes of the two planets becomes the 
natural parameter characterizing coorbital motion, it is useful to
replace variables $(\lambda_1,\Lambda_1;\lambda_2,\Lambda_2)$ by
\begin{eqnarray}
 \theta_1 & \!\!\! = \!\!\! & \lambda_1-\lambda_2\; , \quad J_1 = \frac{1}{2}
  \left(\Lambda_1-\Lambda_2\right)\; , \label{cv1} \\
 \theta_2 & \!\!\! = \!\!\! & \lambda_1+\lambda_2\; , \quad J_2 = \frac{1}{2}
  \left(\Lambda_1+\Lambda_2\right)\; . \label{cv2} 
\end{eqnarray}
The advantage is that $(\theta_1,J_1;\theta_2,J_2)$ remains a set of
canonical variables and $\theta_1$, with $J_1$, are the primary
parameters describing the coorbital motion.

The total Hamiltonian ${\cal H}$, expressed in (\ref{ham1}) and
(\ref{ham2}) as a function of Poincar\'e relative variables, can
be transformed with a lot of algebraic labor into a form depending on
modified Poincar\'e rectangular variables $(\theta_1,J_1;\theta_2,J_2;
x_1,-\imath \bar{x}_1;x_2,-\imath \bar{x}_2)$ (see, e.g., Laskar \&
Robutel 1995; Robutel \& Pousse 2013). In general, ${\cal H} = 
{\cal H}_{\rm K} +{\cal H}_{\rm per} =
{\cal H}_0 + \sum_{k\geq 1}{\cal H}_k$, where ${\cal H}_k\propto x_1^{p_1}
x_2^{p_2}\bar{x}_1^{\bar{p}_1}\bar{x}_2^{\bar{p}_2}$ with positive exponents
such that $p_1+p_2+\bar{p}_1+\bar{p}_2=k$. Hence, ${\cal H}_k$ are
of progressively higher orders in the eccentricities of the two 
planets. We restrict ourselves to the lowest order.

The elegance of the coorbital motion description for small eccentricities
is due to a simple, though rich, form of the fundamental Hamiltonian
${\cal H}_0$. While we shall return to the role of ${\cal H}_1$ and 
higher-order terms in Sec.~2.2, we first discuss the ${\cal H}_0$ term. Note
that ${\cal H}_0$ contains both the Keplerian term ${\cal H}_{\rm K}$
and the fundamental part of the planetary interaction in 
${\cal H}_{\rm per}$.

\subsection{Dynamics corresponding to the ${\cal H}_0$ term}
We find that
\begin{eqnarray}
 {\cal H}_0 & = & -\,G\,\frac{\mu_1 M_1}{2a_1} -G\,\frac{\mu_2 M_2}{2a_2}
  \label{ham3} \\
 & & +\, G m_1 m_2\left[\frac{\cos\theta_1}{\sqrt{a_1 a_2}}-
  \frac{1}{\Gamma\left(a_1,a_2,\theta_1\right)}\right]\; , \nonumber
\end{eqnarray}
where the dependence on the orbital semimajor axes $a_1$ and $a_2$ of the
planets only serves to keep this expression short; the Hamiltonian is
truly a function of the momenta $(J_1,J_2)$ via
\begin{eqnarray}
 a_1 & = & \frac{\left(J_1+J_2\right)^2}{G\mu_1^2 M_1}\; , \label{a11} \\
 a_2 & = & \frac{\left(J_1-J_2\right)^2}{G\mu_2^2 M_2}\; . \label{a21} 
\end{eqnarray}
Additionally, we have
\begin{equation}
 \Gamma\left(a_1,a_2,\theta_1\right) = \sqrt{a_1^2+a_2^2-2a_1a_2
  \cos\theta_1}\; ,  \label{gam}
\end{equation}
which is not to be developed in Taylor series for description of the 
coorbital 
motion at this stage. We also note that a factor $m_0/\sqrt{M_1 M_2}$ 
has been omitted
in the first term of the bracket in Eq.~(\ref{ham3}). This is a
fairly good approximation for planetary masses much smaller than the
stellar mass. We observe that the coordinate $\theta_2$ is absent in 
${\cal H}_0$, implying that the conjugated momentum $J_2$ is constant.
The $J_2$ conservation is just a simpler form of a general angular
momentum integral $2J_2 - |x_1|^2 - |x_2|^2 = C_1$ at this level of 
approximation (eccentricities neglected). The motion is thus reduced
to a single degree of freedom problem ${\cal H}_0 
(\theta_1,J_1;J_2)=C_2$, where $C_2$ is constant. The $C_2$ isolines 
in the $(\theta_1,J_1)$
space provide a qualitative information of system's dynamics.

Further development is driven by observation that in the coorbital
regime $a_1$ and $a_2$ are both very close to some average value
$a_0$. As discussed by Robutel \& Pousse (2013), $a_0$ may
conveniently replace the constant $J_2$ momentum using
\begin{equation}
 J_2 = \frac{1}{2}\left(\mu_1\sqrt{GM_1}+\mu_2\sqrt{GM_2}\right)
  \sqrt{a_0} \; ,  \label{a0}
\end{equation}
In the same time, it is advantageous to introduce a small quantity,
which will characterize small deviation of $a_1$ and 
$a_2$ from $a_0$. This is accomplished by replacing $(\theta_1,J_1)$
with $(\theta,J)$ using a simple shift in momentum:
\begin{equation}
 J_1 = \frac{1}{2}\left(\mu_1\sqrt{GM_1}-\mu_2\sqrt{GM_2}\right)
  \sqrt{a_0} + J \; ,  \label{j}
\end{equation}
and $\theta_1=\theta$. So now ${\cal H}_0={\cal H}_0(\theta,J;a_0)$.
Finally, it is useful to define a dimensionless and small parameter
$u$ instead of $J$ by $J=(\mu_1+\mu_2)\sqrt{Gm_0a_0}\,u$ even at
expense, that $u$ is not canonically conjugated to $\theta$. 
The dynamical evolution of the system is then described by quasi-Hamiltonian
equations
\begin{equation}
 \frac{du}{dt} = -\frac{1}{c}\frac{\partial {\cal H}_0}{\partial \theta}\; ,
  \quad
 \frac{d\theta}{dt} = \frac{1}{c}\frac{\partial {\cal H}_0}{\partial u}\; ,
  \label{ham5} 
\end{equation}
with $c=(\mu_1+\mu_2)\sqrt{Gm_0a_0}$. At this moment it is also useful
to relate $(\theta,u)$ to the semimajor axes of the two planets via
\begin{eqnarray}
 a_1 & = & a_0\left(1+\frac{\mu_1+\mu_2}{\mu_1}\sqrt{\frac{m_0}{M_1}}\,u
  \right)^2 \; , \label{a12} \\
 a_2 & = & a_0\left(1-\frac{\mu_1+\mu_2}{\mu_2}\sqrt{\frac{m_0}{M_2}}\,u
  \right)^2 \; , \label{a22}
\end{eqnarray}
and ${\cal H}_0$ still given by Eq.~(\ref{ham3}).
These relations permit to compute differentiation with respect to
$u$ using the chain rule, such as
\begin{equation}
 \frac{\partial}{\partial u} = \frac{\partial a_1}{\partial u}
  \frac{\partial}{\partial a_1}+\frac{\partial a_2}{\partial u}
  \frac{\partial}{\partial a_2} \; . \label{chr1}
\end{equation}

Once the solution the planet motion in new variables $u(t)$ and $\theta(t)$ 
is obtained, we shall also need to know mean longitudes, $\lambda_1$ and
$\lambda_2$, to determine TTVs. To that end we 
invert Eqs.~(\ref{cv1}) and (\ref{cv2}), obtaining 
\begin{eqnarray}
 \lambda_1 &=& \frac{1}{2}\,\left(\theta_2+\theta\right)\; , \label{lam1} \\
 \lambda_2 &=& \frac{1}{2}\,\left(\theta_2-\theta\right)\; , \label{lam2}
\end{eqnarray}
and find $\theta_2(t)$ from the integration of 
\begin{equation}
 \frac{d\theta_2}{dt} = \frac{\partial {\cal H}_0}{\partial J_2}\; .
  \label{ham6} 
\end{equation}
Differentiation with respect to $J_2$ is obtained by the
chain rule with Eqs.~(\ref{a11}) and (\ref{a21}). 

It is also 
useful to recall that $\theta$ evolves more slowly than $\theta_2$,
since to the lowest order in $u$: $d\theta_2/dt\propto u^0$, while
$d\theta/dt\propto u^1$. In fact, the unperturbed solution reads
$\theta_2\simeq 2\,n_0(t-t_0)$, with 
\begin{equation}
 n_0= \sqrt{\frac{Gm_0}{a_0^3}}\; ,  \label{n0} 
\end{equation}
implying $\lambda_1=\lambda_2\simeq n_0\left(t-t_0\right)$ to the lowest 
order.

Note that so far we considered the exact solution of ${\cal H}_0$, without 
referring  to approximations given by its expansion in small quantities: 
$u$, and $m_1/m_0$ and $m_2/m_0$. The 
reason for this was twofold. First, we found such series may converge slowly and 
truncations could degrade accuracy of the solution. Second, although we 
find it useful to discuss some aspects of such development in the small 
parameters below, we note that the system is not integrable analytically
at any meaningful approximation. This implies that semi-numerical approach is
anyway inevitable. Considering the complete system, as opposed to
approximations given by truncation of series in the above mentioned small
parameters, does not extent the CPU requirements importantly. In fact,
Eqs.~(\ref{ham5}) and (\ref{ham6}) are easily integrated by numerical 
methods (in our examples below we used simple Burlish-Stoer integrator
leaving implementation of more efficient symplectic methods for future 
work). 
\begin{figure}[t]
 \epsscale{1.}
  \plotone{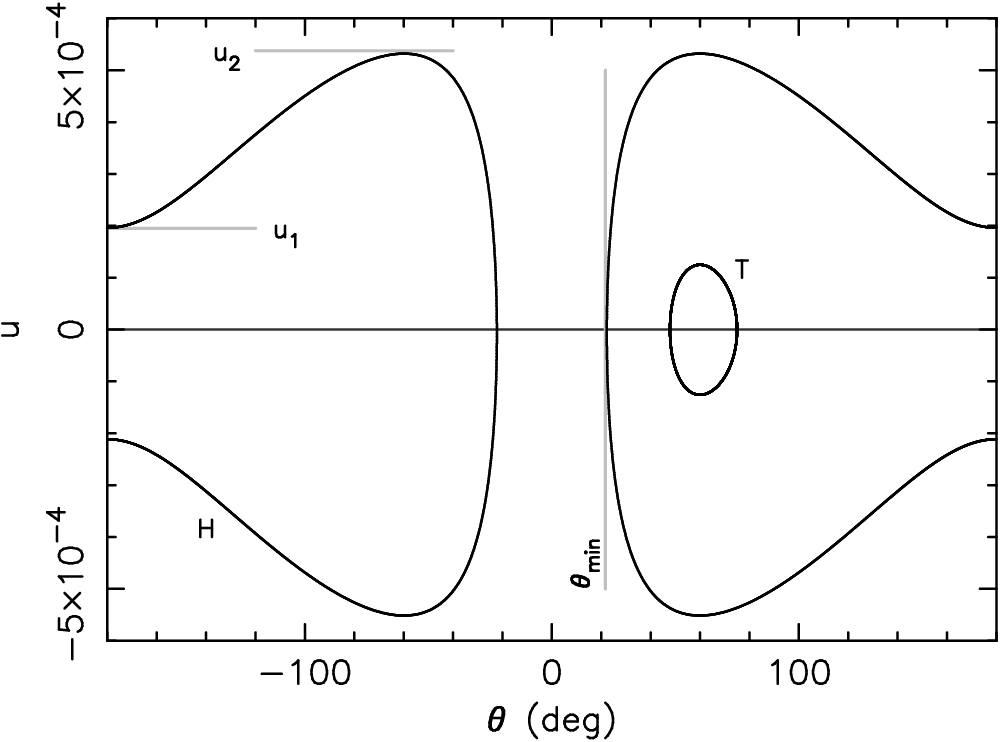}
 \caption{Example of two trajectories in the $(u,\theta)$ phase space of
   Hamiltonian ${\cal H}_0$; for sake of illustration we used $m_0=1$~M$_\odot$,
   $m_1/m_0=10^{-3}$ and $m_2/m_0=2\times 10^{-5}$. The trajectory in a horseshoe
   regime, labeled H, is characterized by: (i) minimum and maximum value of
   $|u|$ parameter (denoted here $u_1$ and $u_2$), and (ii) minimum angular
   separation $\theta_{\rm min}$ of the coorbiting planets. The trajectory
   labeled T shows an example of a tadpole orbit, librating about the
   Lagrangian stationary point L4 for comparison.}
 \label{f1}
\end{figure} 

While numerical approach provides an exact solution, it is still useful
to discuss some qualitative aspects by using the approximated forms of
${\cal H}_0$. The smallness of $u$ permits denominator factors 
such as $1/\Gamma$ in Eq.~(\ref{ham3}) be developed in power series of 
which we preserve terms up to the second order (see also Robutel \& Pousse 2013):
\begin{equation}
 {\cal H}_0\left(\theta,u;a_0\right) =\frac{G}{a_0}\left( A_0 +
   A_1\,u + A_2\,u^2\right) \; .  \label{ham4}
\end{equation}
The $A$-coefficients read
\begin{eqnarray}
 A_0 & \!\!\! = \!\!\! & \frac{\sigma}{2}\left(2-\gamma^2-
  \frac{2}{\gamma}\right)\; , \label{ca0} \\
 A_1 & \!\!\! = \!\!\! & -\frac{\sigma_+\sigma_-}{2}\left(1-\gamma\right)^2
 \left(1+\frac{2}{\gamma}\right)\; , \label{ca1} \\
 A_2 & \!\!\! = \!\!\! & -\frac{3}{2}\frac{m_0\sigma_+^3}{\sigma}+
  \label{ca2} \\
  & \!\!\!  \!\!\! & \frac{\sigma_+^2}{\sigma}\left[\left(\sigma_+^2-3\sigma
  \right)\left(4-\frac{\gamma^2}{2}-\frac{1}{\gamma}\right)+
  \frac{2\sigma_+^2}{\gamma^3}\right] \; , \nonumber
\end{eqnarray}
with the mass-dependent parameters $\sigma_\pm = m_1\pm m_2$ and 
$\sigma=m_1m_2$, and $\gamma = \sqrt{2-2\cos\theta}$. Terms to the
second power of planetary masses have been retained in Eq.~(\ref{ham4}).
In fact, the simplest form is obtained by dropping the linear term in
$u$, and approximating $A_2$ by the first factor only.%
\footnote{Note that the second term in $A_2$ is by a factor $\propto
 \sigma_+/m_0$ smaller than the first term. The smallness of the term linear
 in $u$ is obvious in the limit of planets with a similar mass for
 which $\sigma_-\simeq 0$. In the regime of planets with unequal masses,
 $m_2\ll m_1$, one finds that $A_1 u/A_0 \propto (m_1/m_0)^{1/3}$,
 rendering the omitted linear term again small (see Eqs.~(27) and (28)
 in Robutel \& Pousse 2013).}
This results in
\begin{equation}
 {\cal H}_0 = -\frac{3}{2}\frac{G}{a_0}\frac{m_0\sigma_+^3}{\sigma} u^2
  + \frac{G\sigma}{a_0}\left(\cos\theta - \frac{1}{\gamma}\right) \; , 
 \label{ham7}
\end{equation}
introduced already by Yoder et~al. (1983) (see also Sicardy \& Dubois
2003). Hamiltonian (\ref{ham7}) corresponds to a motion of a particle
in the potential well
\begin{equation}
 U\left(\theta\right) = \cos\theta - \frac{1}{\gamma(\theta)}\; .
  \label{effpot}
\end{equation}
As discussed by Robutel \& Pousse (2013), in both approximations (\ref{ham4})
and (\ref{ham7}) the exact character of motion is not represented near
$\Gamma\simeq 0$, corresponding to a collision configuration, but this is not 
of great importance for us.

Unfortunately, the Hamiltonian (\ref{ham7}) is not integrable analytically.
Still, the energy conservation ${\cal H}_0=C_2$ provides a qualitative
insight into trajectories in $(u,\theta)$ space and also
allows us to quantitatively estimate some important parameters. Figure~\ref{f1}
shows examples of two trajectories in phase space of $(u,\theta)$, one corresponding
to a horseshoe solution (H) and one corresponding to a tadpole solution (T)
librating about the L4 Lagrangian stationary solution. Here, we set $m_0$ equal
to solar mass, $m_1/m_0=10^{-3}$ and $m_2/m_0=2\times 10^{-5}$.
Since we are primarily focusing on the horseshoe coorbital regime, we
determine relations between parameters characterizing the 
H-trajectory in Fig.~\ref{f1}. These are the: (i) minimum $u_1$ and maximum $u_2$ 
amplitudes of $|u|$ along the trajectory, (ii) minimum separation angle
$\theta_{\rm min}$, and (iii) half-period $T$ of motion along the trajectory
in the $(u,\theta)$ space. One easily finds that $u_1$ corresponds to
planetary opposition $\theta=\pm \pi$, and $u_2$ corresponds to longitude 
of the Lagrangian stationary solutions $\theta=\pm \pi/3$. As a result
\begin{equation}
 u_2^2-u_1^2 = \frac{2}{3}\frac{\sigma^2}{m_0\sigma_+^3}\; . \label{u1u2}
\end{equation}
The symmetry of ${\cal H}_0$ in $u$ implies that $\theta_{\rm min}$
corresponds to $u=0$, and thus
\begin{equation}
 u_1^2 = \frac{1}{3} \frac{\sigma^2}{m_0\sigma_+^3}\left(2\,\Sigma_{\rm min}+
  \sqrt{\frac{2}{\Sigma_{\rm min}}}-5\right)\; , \label{u1th}
\end{equation}
where we denoted $\Sigma_{\rm min}=1-\cos\theta_{\rm min}$. The inverse
relation requires solution of a cubic equation, conveniently given
in the standard form. By using the trigonometric formulas one has
\begin{equation}
 \Sigma_{\rm min} = \frac{4\,K}{3}\cos^2\left[\frac{1}{3}\,{\rm acos}
  \left(\sqrt{\frac{27}{8\,K^3}}\right)-\frac{2\pi}{3}\right]\; ,
 \label{thu1}
\end{equation}
with
\begin{equation}
 K = \frac{3}{2}\frac{m_0\sigma_+^3}{\sigma^2}\, u_1^2+\frac{5}{2}
  \; . \label{c2small}
\end{equation}
The critical trajectory, representing transition between the horseshoe and 
tadpole orbits, has $u_1=0$, thus $K=5/2$. Equation (\ref{thu1}) then
provides a formula for a maximum value of $\theta_{\rm min}$ separation
in the horseshoe regime, roughly $23.9^\circ$. The minimum value of
$\theta_{\rm min}$ is approximately set by Lagrangian $L_1$ and $L_2$ stationary
points of the ${\cal H}_0$ Hamiltonian. Robutel \& Pousse (2013) show
that this minimum separation value is $\simeq \case45 (\sigma_+/6m_0)^{1/3}$.
Depending on planetary masses defining $\sigma_+/m_0$ this may be few
degrees.
\begin{figure}[t]
 \epsscale{1.}
  \plotone{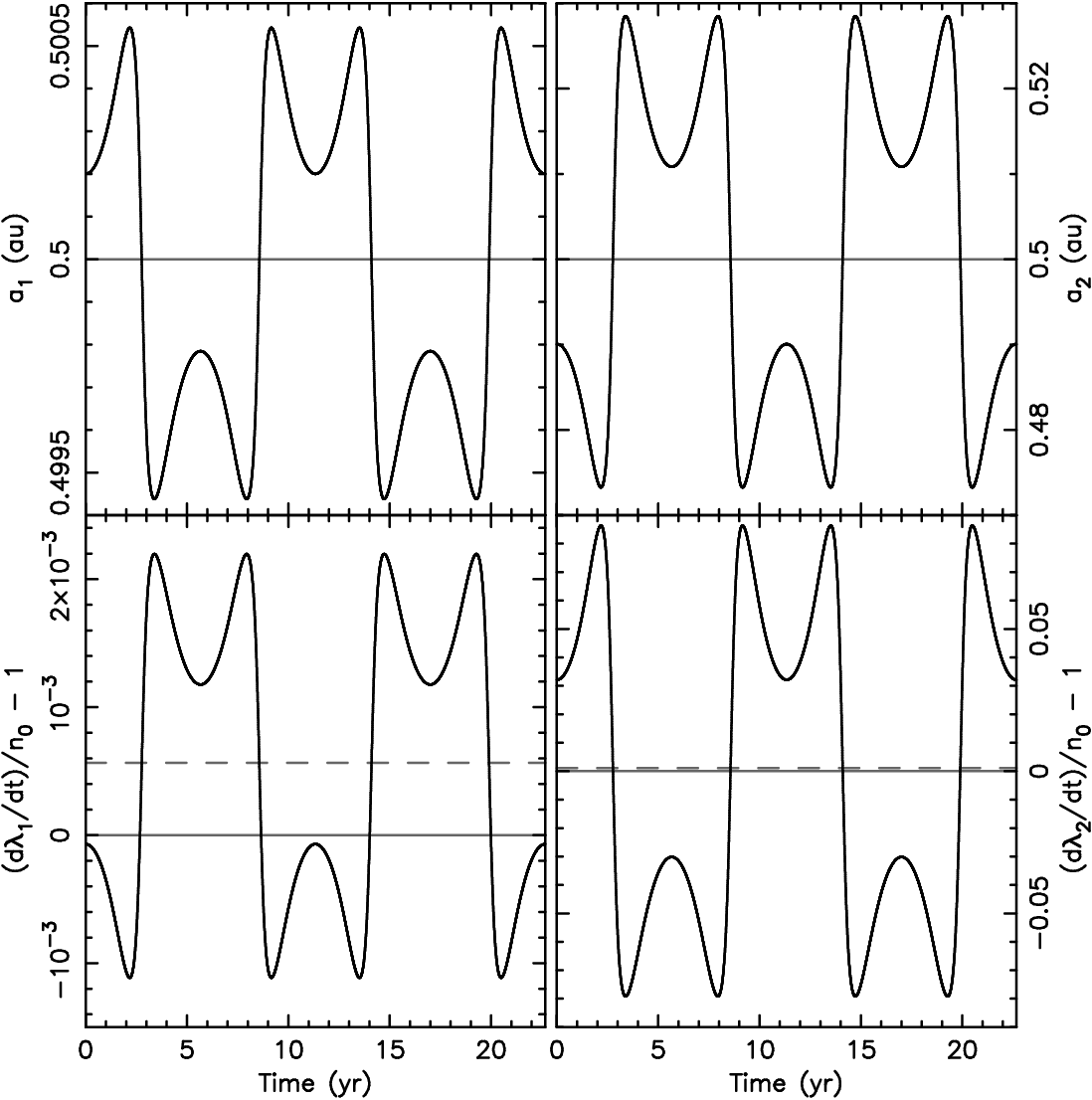}
 \caption{Top: Semimajor axis $a_1$ of the heavier planet (left) and the
   lighter planet $a_2$ (right) as a function of time for the
   horseshoe trajectory in Fig.~\ref{f1}. Their mean value $a_0=0.5$~au, for
   definiteness. The timespan covers two cycles of the circulation along
   the trajectory. The lighter planet experiences
   larger perturbations and thus larger-amplitude variations of the semimajor
   axis. Bottom: Normalized rates of change
   of the longitude in orbit $(d\lambda_1/dt)/n_0-1$ of the heavier planet
   (left) and the lighter planet $(d\lambda_2/dt)/n_0-1$ (right). The
   dashed horizontal lines are numerically computed mean values. They
   are offset from zero for reasons discussed in the main text.}
 \label{f2}
\end{figure} 

Finally, the relation between $T$ and $\theta_{\rm min}$ is
obtained from the energy conservation%
\footnote{Equation~(\ref{t}) can be readily obtained by expressing $u$ using
 $d\theta/dt$ from the second of the Hamilton equations (\ref{ham5}),
 and plugging it in the Hamiltonian (\ref{ham7}).}:
\begin{equation}
 \sqrt{\frac{3}{2}\frac{\sigma_+}{m_0}}\, n_0 T = \int_{\theta_{\rm min}}^{\pi} 
  \frac{d\theta}{\sqrt{U(\theta)-c_2}}\; , \label{t}
\end{equation}
where $c_2 = C_2 (a_0/G\sigma)=U(\theta_{\rm min})=1-K$, with $K$ given
above. Obviously, $a_0$ is now needed to  
gauge $n_0$, while $c_2$ may be obtained as a function of either of the 
parameters $u_1$, $u_2$ or $\theta_{\rm min}$( see Eq.~(\ref{c2small})).
If one wants the right hand side in (\ref{t}) be solely a function of
$\theta_{\rm min}$, we have
\begin{equation}
 c_2 = 1-\left(\Sigma_{\rm min}+ \frac{1}{\sqrt{2\,\Sigma_{\rm min}}}\right)\; .
  \label{c2smallbis}
\end{equation}

Figure~\ref{f2} shows time evolution of the semimajor axes $a_1$ and 
$a_2$, and mean orbital longitude rates for the exemplary system
shown in Fig.~\ref{f1} ($m_0=1$~M$_\odot$, $m_1/m_0=10^{-3}$, $m_2/m_0
=2\times 10^{-5}$, and $a_0=0.5$~au). In this case we focus on the horseshoe 
orbit denoted H on Fig.~\ref{f1}. The longitude rates, computed from
their definition (\ref{lam1}) and (\ref{lam2}) and using
Eqs.~(\ref{ham5}) and (\ref{ham6}), are 
represented in a normalized way by $(d\lambda_1/dt)/n_0-1$ and
$(d\lambda_2/dt)/n_0-1$.  As the trajectory moves along the oval-shaped
curve in the phase space, the orbits periodically switch their positions
with respect to the star causing their semimajor axis to jump around the
$a_0$ value. Each orbit stays at higher/lower-$a$ regime for time $T$,
which is approximately $\propto \sqrt{m_0/\sigma_+}/3$ longer than its
orbital period. The switch in semimajor axes is reflected in the
corresponding variations in longitude rate. Note
that the average $\lambda$-rates for both orbits are not equal to
$n_0$ from (\ref{n0}), producing an offset shown by the difference of the
dashed line and zero at the bottom panels of Fig.~\ref{f2}. This is
because of the planetary masses also contributing to their mean 
motion about the star, while the definition of $n_0$ as a nominal,
reference fast frequency did not take this into account. In fact, we find
that the longitude rate normalized by $n_0$ and averaged over the
coorbital cycle $2\,T$ is approximately $1+(m_i/2m_0)+ 6\,(\sigma_+/m_i)^2
u_\star^2+\ldots$ for each of the orbits $(i=1,2)$, with $u_\star$ being
a characteristic value of the $u$-parameter over one of the half-cycles
(e.g., one could approximate $u_\star\simeq (u_1+u_2)/2$). When the 
planetary masses are unequal, as in our example, the more massive planet
has the mean longitude rate primarily modified by its own mass.
In our case $(d\lambda_1/dt)/n_0-1 \simeq m_1/(2m_0)$. The lighter 
planet's longitude rate is dominated by the second term, i.e., in our
case $(d\lambda_2/dt)/n_0-1 \simeq 6\, (\sigma_+/m_2)^2 u_\star^2$.

Even more important for the TTV analysis is to consider how much the
mean rates in longitude change during the coorbital
cycle. We find that the change in longitude rates during the
low/high-$a$ states for each of the planets is approximated by
$\simeq \pm 6\,n_0\, (\sigma_+/m_i)\, u_\star$ $(i=1,2)$. If large enough,
this value may build over a timescale $T$ to produce large variations in the
mean longitude of planets, and thus result in large TTVs. We
will discuss this in Sec.~3.

\subsection{Eccentricity terms}
So far, we approximated the interaction Hamiltonian ${\cal H}$ with the
leading part ${\cal H}_0$ from Eq.~(\ref{ham3}) that is independent of 
eccentricities $e_1$ and $e_2$. In order to extend our analysis to
the regime of small-$e$ values, we include the lowest-order interaction contributions.
Neglecting the second-order eccentricity terms, we have ($z=e\,\exp(\imath \varpi)$)
\begin{equation}
 \frac{dz}{dt} = -\frac{2\imath}{\Lambda}\,\frac{\partial {\cal H}}{\partial
  \bar{z}} \; , \label{hamz}
\end{equation}
where we insert ${\cal H}={\cal H}_1$ and secular part from ${\cal H}=
{\cal H}_2$, the first- and second-order terms in eccentricity development of
${\cal H}$. When $\delta z(t) = z(t)-z(0)$ is known for both orbits from 
solution of (\ref{hamz}), we can compute their effect on TTVs by defining 
(e.g., Nesvorn\'y \&  Vokrouhlick\'y 2014) 
\begin{equation}
 \delta \lambda^{\rm eff} = \imath\left(\delta z\, e^{-\imath \lambda_0} - 
  \delta \bar{z}\, e^{\imath \lambda_0} \right)\; . \label{lameff}
\end{equation}
Here $\lambda_0$ is the unperturbed longitude in orbit for which we substitute
the zero-order solution $\lambda_0=n_0(t-t_0)$ plus a phase, individual to each
of the two planets. This is an effective change in orbital longitude given
here to the first order in eccentricity (see Nesvorn\'y 2009 for higher order 
terms), which together with the direct
effect in $\lambda$ contributes to TTVs. It is not known a priori which of
these contributions should be more important. For instance, in the case of
closely packed (but not coorbiting) orbits studied by Nesvorn\'y \&
Vokrouhlick\'y (2014), the eccentricity term (\ref{lameff}) was generally
larger than the direct perturbation in $\lambda$ over a short-term timescale.

We should also note that the ${\cal H}_1$ and 
${\cal H}_2$ Hamiltonians would also contribute to variations of the
$(u,\theta,\theta_2)$ variables. Perhaps the most interesting effect should
be a slight modification of the planetary mean motion through the change
in $\theta_2(t)$. However, since it is not our intention to develop a complete
perturbation theory for coorbital motion here, we neglect these terms
focusing on the lowest-order eccentricity effects. We verified that a
slight change in initial conditions, specifically the $u$ parameter value,
would equivalently represent the eccentricity modification of the 
$\theta_2(t)$ angle.

\subsubsection{First-order terms}
We start with the first-order eccentricity terms in ${\cal H}_1$. While
apparently of a larger magnitude in ${\cal H}$ than ${\cal H}_2$, they are 
short-periodic and this diminishes their importance. An easy algebra shows
that the perturbation equations read (recall that the overbar means complex
conjugation)
\begin{eqnarray}
 \frac{dz_1}{dt} & \!\!=\!\! & -n_0\,\frac{m_2}{m_0}\,\Phi\left(\theta\right)\,
  e^{\imath \theta_2/2}\; , \label{zper1} \\
 \frac{dz_2}{dt} &\!\! =\!\! & \phantom{-}n_0\,\frac{m_1}{m_0}\,{\bar \Phi}\left(\theta
  \right)\, e^{\imath \theta_2/2}\; , \label{zper2} 
\end{eqnarray}
with
\begin{equation}
 \Phi\left(\theta\right) = \imath\,e^{3\imath \theta/2}+\frac{\sin\frac{1}{2}
  \theta}{\gamma^3}\left(3+e^{\imath \theta}\right) \; .  \label{phif}
\end{equation}
We neglected terms of the order $u$ and higher in the right hand sides of
(\ref{zper1}) and (\ref{zper2}), and used
$\gamma=\gamma(\theta)=\sqrt{2-2\cos\theta}$. Since $\theta_2/2\simeq
n_0(t-t_0)$, the power-spectrum of the right hand sides in (\ref{zper1}) and 
(\ref{zper2}) is indeed dominated by the high (orbital) frequency $n_0$,
modulated by slower terms from $\Phi$ dependence on $\theta$.

\subsubsection{Second-order terms}
The second-order eccentricity terms in ${\cal H}_2$ are important, because
they are the first in the higher-order ${\cal H}$ expansion part to depend 
on low-frequencies only. Restricting to this part of ${\cal H}_2$, thus dropping
the high-frequency component in ${\cal H}_2$, we obtain (see Robutel \& Pousse 
2013)
\begin{eqnarray}
 \frac{dz_1}{dt} & \!\!=\!\! & -2\imath\,n_0\,\frac{m_2}{m_0} \left(A\, z_1 + B\,
   z_2\right)\; , \label{zper3} \\
 \frac{dz_2}{dt} & \!\!=\!\! & -2\imath\,n_0\,\frac{m_1}{m_0} \left({\bar B}\,
   z_1 + A\, z_2\right)\; , \label{zper4} 
\end{eqnarray}
with
\begin{eqnarray}
 A & \!\!\! = \!\!\! & \frac{1}{8\gamma^5}\left(5\cos 2\theta - 13\right)
  - \frac{\cos\theta}{2}\left(1-\frac{1}{\gamma^5}\right)\; , \nonumber \\
   & \!\!\!  \!\!\! & \label{af} \\
 B & \!\!\! = \!\!\! & \frac{1}{2}\left(1-\frac{2}{\gamma^5}\right)
  e^{2\imath \theta} + \label{bf} \\
   & \!\!\!  \!\!\! & \frac{1}{8\gamma^5}\left[\imath\sin\theta\left(9-
  e^{2\imath \theta}\right)+8e^{\imath \theta}\right]\; . \nonumber
\end{eqnarray}
We again neglected terms proportional to $u$ and its powers in
expressions for $A$ and $B$ for simplicity.
While the right hand sides of Eqs.~(\ref{zper3}) and (\ref{zper4}) are of
the first order in eccentricities $e_1$ and $e_2$, they do not contain
high-frequency terms and thus the corresponding perturbations may accumulate 
over time to large values. Indeed, these are the secular perturbations
dominating the eccentricity changes.

The equations (\ref{zper1}) -- (\ref{zper4}) do not possess
analytical solutions. Therefore,
we numerically integrated them together with those for $u$, $\theta$
and $\theta_2$, to determine $z_1(t)$ and $z_2(t)$.

\section{An exemplary case}
We now give an example of a coorbital system about a solar
mass star and compute TTVs by two methods: direct numerical integration
of the system in Poincar\'e relative variables and using the theory
presented in Sec.~2. 

We used the same planetary configuration whose short-term dynamics was 
presented in Figs.~\ref{f1} and \ref{f2}. In particular, 
$m_0=1$~M$_\odot$ star with a Jupiter-mass planet $m_1=10^{-3}$~M$_\odot$
coorbiting with a sub-Neptune mass planet $m_2=2\times 10^{-5}$~M$_\odot$.
The mean distance from the star was set to be $a_0=0.5$~au. The initial orbits
were given small eccentricities of $e_1=e_2=0.01$, and colinear
pericenter longitudes $\varpi_1=\varpi_2=0^\circ$. The initial longitude in
orbit of both planets were $\lambda_1=180^\circ$ and $\lambda_2=0^\circ$,
such that at time zero they were at opposition. 

Starting with these
initial data, we first numerically integrated the motion using
Poincar\'e relative coordinates $({\bf r}_1,{\bf r}_2; {\bf p}_1,{\bf p}_2)$
introduced in Sec.~1. The equations of motion were obtained from
the Hamiltonian ${\cal H}={\cal H}_{\rm K}+{\cal H}_{\rm per}$, with the
two parts given by Eqs.~(\ref{ham1}) and (\ref{ham2}). For our simple
test we used a general purpose Burlish-Stoer integrator with a tight
accuracy control. The integration timespan was $\simeq 22.7$~yr
covering two cycles of the coorbital motion (see Fig.~\ref{f2}). For
sake of definiteness, we assumed an observer along the x-axis of the
coordinate system and we numerically recorded times of transit of the
two planets. The transit timing variations were obtained by removing linear
ephemeris from transits.

Next, we assumed the system is described by a set of parameters $(u,\theta,
\theta_2; z_1,z_2)$ introduced and discussed in Sec.~2, and numerically 
integrated their dynamical equations (\ref{ham5}), (\ref{ham6}), and
(33-39). For each of the planets we then
computed TTVs from (e.g., Nesvorn\'y \& Morbidelli 2008, Nesvorn\'y \&
Vokrouhlick\'y 2014)
\begin{equation}
 -n_\star \,\delta t = \delta \lambda + \delta \lambda^{\rm eff} \; ,
  \label{ttv}
\end{equation}
where $n_\star$ is the effective mean motion of the unperturbed motion.
We use the mean values of the longitude in orbit rate discussed in
Sec.~2.1, for instance $n_\star=n_0\,(1+m_1/2m_0)$ for the Jupiter-mass
planet. Having $\theta(t)$ and $\theta_2(t)$ integrated, we recover the
time-dependence of the longitudes $\lambda_1(t)$ and $\lambda_2(t)$
from (\ref{lam1}) and (\ref{lam2}). From these numerically-determined
functions we subtracted the
average mean motion trend $n_\star \,(t-t_0)$ and obtained variation
$\delta \lambda$ of both planets as needed for the computation of TTVs
(Eq.~\ref{ttv}). The effective eccentricity terms $\delta \lambda^{\rm eff}$
were computed from their definition in Eq.~(\ref{lameff}).
\begin{figure}[t]
 \epsscale{1.}
  \plotone{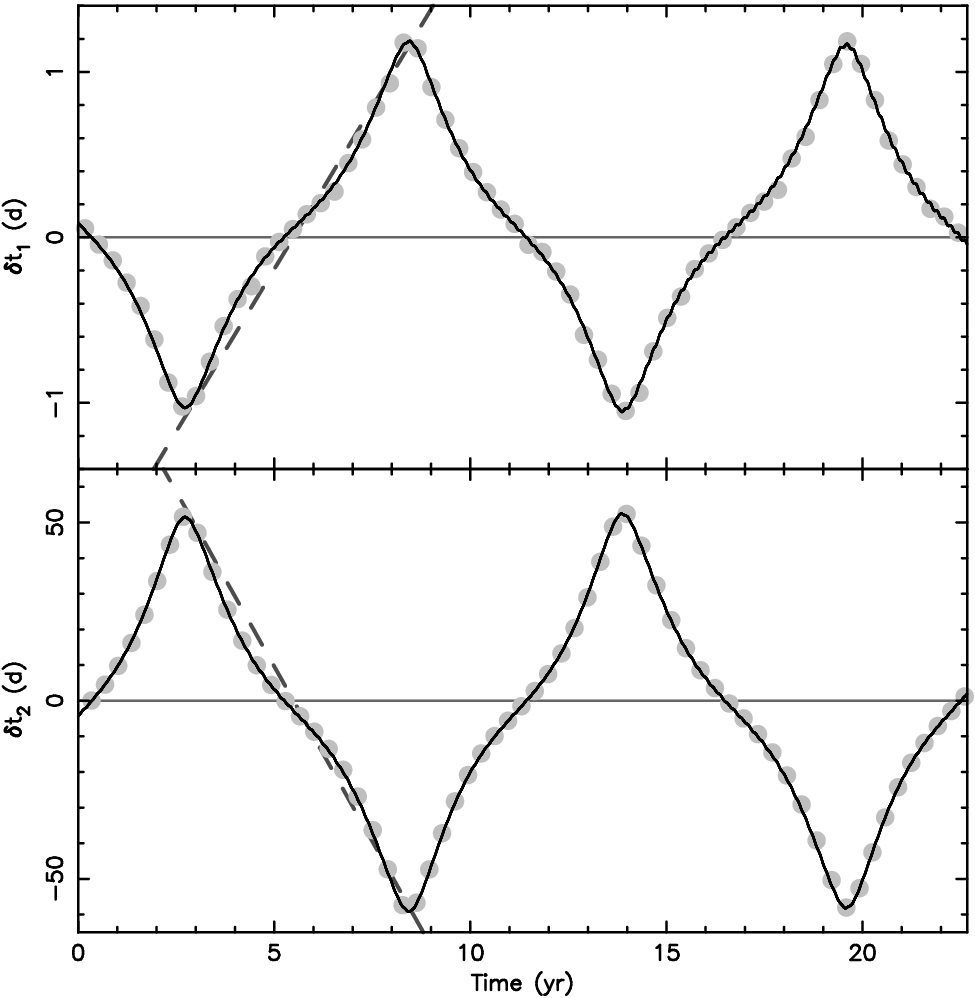}
 \caption{Transit timing variations (TTVs) of the heavier planet $\delta t_1$
   (top) and the lighter planet $\delta t_2$ (bottom). Planetary masses and 
   mean semimajor axis value as in Figs.~\ref{f1} and \ref{f2}. The
   gray symbols are the TTVs obtained from direct numerical integration
   in Poincar\'e relative variables (Eqs.~\ref{ham1} and
   \ref{ham2}). The solid line is from the semi-analytic theory given by
   Eq.~(\ref{ttv}). The dashed and sloped lines in both panels 
   are from our expected amplitude of change in longitude rate during the
   switches between legs in the coorbital cycle; the slope estimate is
   $\simeq 3\,(\sigma_+/m_i) u_\star$, with $u_\star\simeq 3.55\times 10^{-4}$
   and planets $i=1$ (top) and $i=2$ (bottom). See the main text for more details.}
 \label{f3}
\end{figure} 
\begin{figure}[t]
 \epsscale{1.}
  \plotone{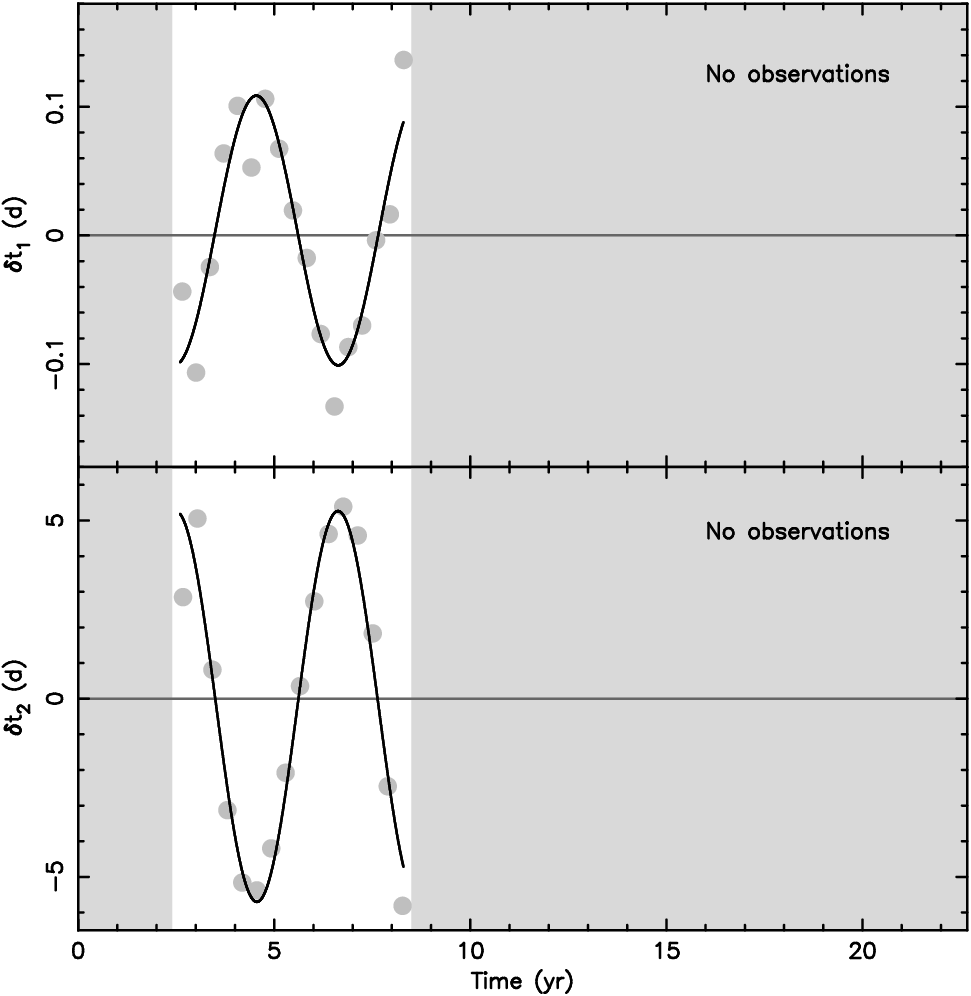}
 \caption{Transit timing variations (TTVs) of the heavier planet $\delta t_1$
   (top) and the lighter planet $\delta t_2$ (bottom) from Fig.~\ref{f3} if
   the dataset is limited to observations between 2.6~yr and 8.3~yr. In this
   case, the analysis would reveal misleading signal with quasi-periodicity
   of $\simeq 4.2$~yr and much smaller amplitude (solid lines).}
 \label{f4}
\end{figure} 

Figure~\ref{f3} shows a comparison between the synthetic TTVs from direct
numerical integration (gray symbols) compared to the $\delta t$ function
from Eq.~(\ref{ttv}) (black line). For sake of the example we assumed an
ideal situation with both planets transiting. As mentioned in Sec.~2.2,
we used a small change in the initial conditions of the secular theory,
namely rescaled the $u$ parameter by fractionally $\propto e^2$ value, to
represent the ${\cal H}_2$ effect on the mean motion of planets. With
that adjustment, the match between the synthetic TTV series and the
modeled function $\delta t$ is excellent. We also note that the contribution
of the second term in the right-hand side of (\ref{ttv}) is negligible and
basically all effect seen on the scale of Fig.~\ref{f3} is due to the
first term (i.e., direct perturbation in orbital longitude). The dashed sloped
lines on both panels of Fig.~\ref{f3} show the effect of a change in mean
motion of the planets, as estimated from the simple Hamiltonian (\ref{ham7}).
In particular their slopes are: (i) $3\,(\sigma_+/m_1)\,u_\star$ at the top panel, 
and (ii) $-3\,(\sigma_+/m_2)\,u_\star$ at the bottom panel ($u_\star
\simeq 3.55\times 10^{-4}$). The match to the mean behavior of the TTVs is
good, since in the simplest approximation the planets motion may be understood
as a periodic switch between two nearly circular orbits. Since the period $T$
is about 15 times longer than the orbital period of the planets in our case,
the effect may accumulate into a large amplitude TTV series.
\begin{figure}[t]
 \epsscale{1.}
  \plotone{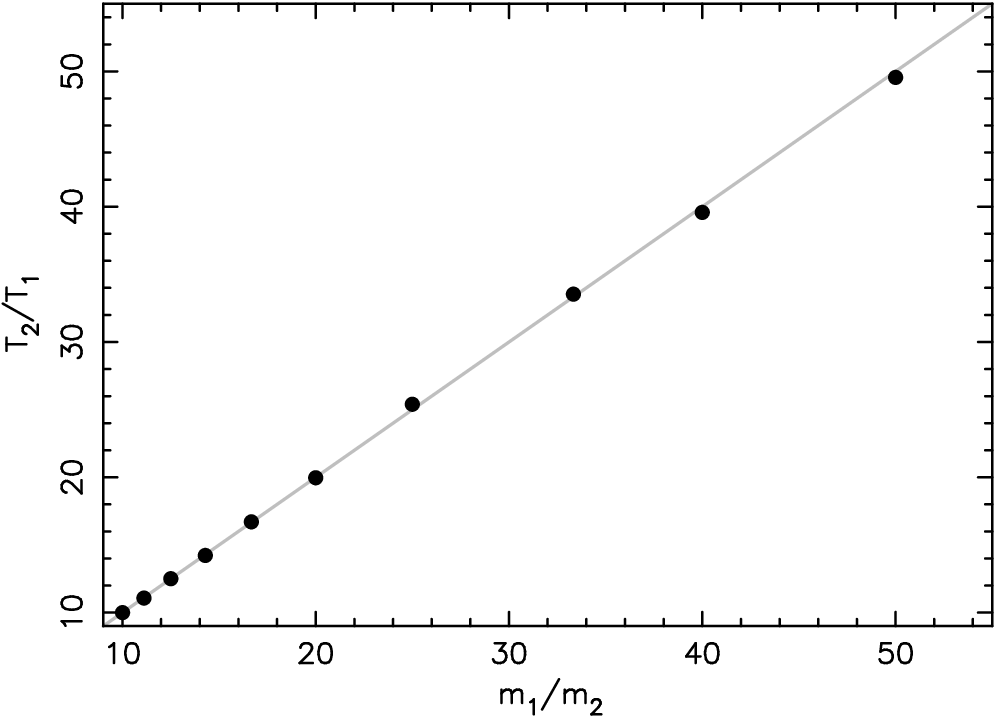}
 \caption{Ratio $T_2/T_1$ of the maximum amplitude of the TTV series for the
 less massive planet ($T_2$) vs those of the more massive planet ($T_1$) 
 placed onto a horseshoe orbit in the 1:1 MMR as determined from the full-fledged
 numerical integration (ordinate). The abscissa is the planetary mass ratio
 $m_1/m_2$. The gray line if the direct proportionality rule obtained from the
 simplified analytical theory (Sec.~2.1). Details of the set-up are described in
 the main text.}
 \label{f5}
\end{figure} 

It is interesting to point out that recognizing the planet's configuration
requires observations covering at least the fundamental
period $T$ of the coorbital motion. For instance, if the observations would have
covered a shorter interval, say between 2.6~yr and 8.3~yr in Fig.~\ref{f3}, one
may not recognize the coorbital signature in the TTVs. Figure~\ref{f4}
shows how the data would have looked in this case. TTV series of both planets
would look quasi-periodic with a period $\sim T/2\simeq 4.2$~yr, reflecting
behavior of the planets' mean motion variation over one quarter of the
coorbital cycle (i.e., when $\theta$ leaps from $\theta_{\rm min}$ to
$360^\circ-\theta_{\rm min}$, Fig.~\ref{f1}). Equation~(\ref{t}) indicates
that $T\propto a_0^{3/2}\,(m_0/\sigma_+)^{1/2}$, making thus the necessary
observational timescale (i) shorter for closer-in planets, and (ii) longer
for less massive planets. So for instance $\simeq 12$~y periodicity of the TTV 
series shown in Fig.~\ref{f3} would also hold for about $8$~Earth mass
coorbiting planets at about $0.15$~au distance (i.e., $\simeq 20$~d revolution
period) from a solar mass star. These are very typical systems observed by
the Kepler satellite.

Consider now an ideal situation when both planets are transiting and a
long enough series of TTVs are recorded for both of them (e.g., Fig.~\ref{f3}).
Analysis based on approximate Hamiltonian (\ref{ham7}) then suggests (Sec.~2.1) 
that the ratio of maximum amplitudes
of the TTV series, to be denoted $T_1$ for the more massive planet and $T_2$ for
the less massive planet, is equal to the ratio of their masses: $T_2/T_1=
m_1/m_2$. Since $T_2/T_1$ can be measured from the observations, the mass ratio
of the coorbital planets is readily constrained. To verify validity
of this conclusion, we numerically integrated a complete Hamiltonian in
Poincar\'e rectangular coordinates with a solar mass star having two
coorbital planets with masses $m_1=10^{-3}$~M$_\odot$ and $m_2$ ranging values
$2\times 10^{-5}$~M$_\odot$ to $10^{-4}$~M$_\odot$. We used $a_0=0.5$~au
and set the planets initially at opposition, i.e. giving them $\lambda_1=180^\circ$ 
and $\lambda_2=0^\circ$. The initial eccentricity values were assumed small,
$e_1=e_2=0.001$, and pericenter longitudes $\varpi_1=\varpi_2=0^\circ$.
For each of the mass configurations considered, we followed the system for 1000~yr
and derived the synthetic TTV series as shown by symbols in
Fig.~\ref{f3}. We then fitted the maximum amplitudes $T_1$ and $T_2$. Their
ratio is shown by black circles in Fig.~\ref{f5}, while the gray line is
the expected direct proportionality relation mentioned above. We note the
linear trend is a very good approximation.
\begin{figure}[t]
 \epsscale{1.}
  \plotone{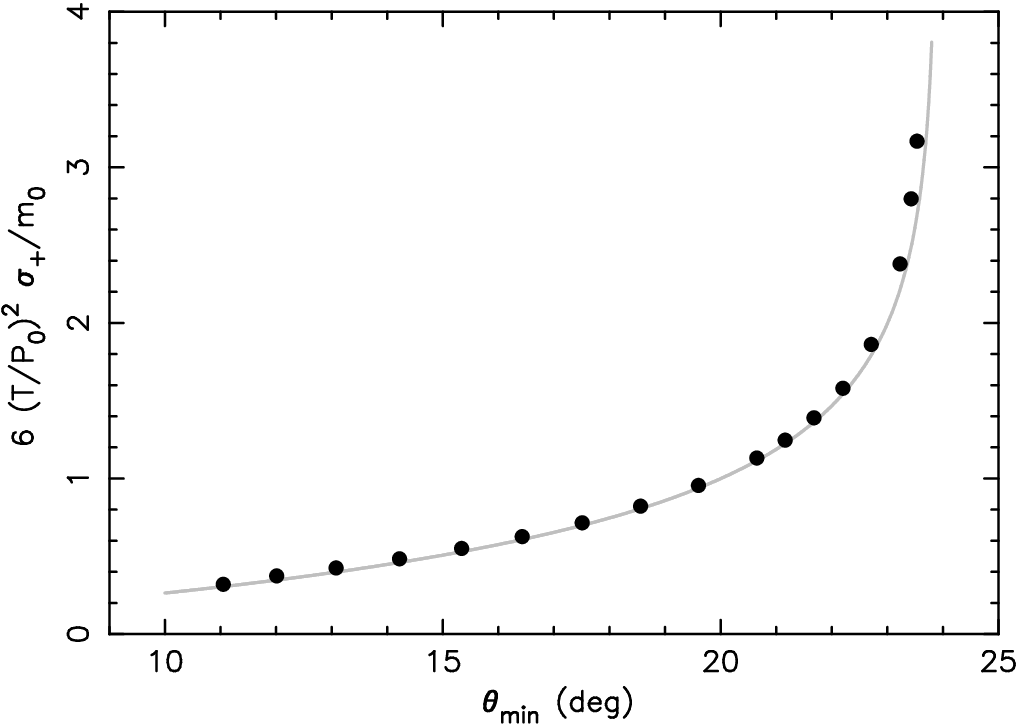}
 \caption{Correlation between the $6\,(T/P_0)^2\,(\sigma_+/m_0)$ parameter
  (ordinate) and the minimum angular separation $\theta_{\rm min}$ of the coorbiting
  planets (abscissa). Here $P_0$ is the mean orbital period of the planets, $T$
  is the half period of the coorbiting cycle, $\sigma_+=m_1+m_2$ is the total mass
  of the planets and $m_0$ is the stellar mass. The black symbols are from
  direct numerical integration. The gray line
  is the ${\cal J}(\theta_{\rm min})$ function from  Eq.~(\ref{jint}), suggested
  from a simple analytic theory. Beyond $\theta_{\rm min}\simeq 25^\circ$ the
  orbital configuration transits to the tadpole regime.}
 \label{f6}
\end{figure} 

Another useful parametric constraint is hinted by Eq.~(\ref{t}), again obtained
from the simplified Hamiltonian form (\ref{ham7}). Denoting for short
\begin{equation}
 {\cal J}\left(\theta_{\rm min}\right)= \int_{\theta_{\rm min}}^{\pi} 
  \frac{d\theta}{\sqrt{U(\theta)-c_2}}\; , \label{jint}
\end{equation}
solely a function of the minimum angular separation $\theta_{\rm min}$ of the
planets, we have
\begin{equation}
 6\,\left(\frac{T}{P_0}\right)^2\frac{\sigma_+}{m_0} = \frac{1}{\pi^2}\,
  {\cal J}^2\left(\theta_{\rm min}\right)\; . \label{link2}
\end{equation}
Here $P_0=2\pi/n_0$ is a good proxy for the mean orbital period of the planets,
$T$ is the half period of the coorbiting cycle (i.e., time between minima and
maxima of the TTVs, Fig.~\ref{f3}), and $\sigma_+=m_1+m_2$ as above. 
Since $T/P_0$ can be directly constrained from the observations,
Eq.~(\ref{link2}) provides a link between the mass factor $\sigma_+/m_0$
and $\theta_{\rm min}$. We tested the validity of Eq.~(\ref{link2}) by directly
by integrating the planetary system in Poincar\'e rectangular coordinates.
The model parameters were mostly the same as above, except for: (i) fixing now the
planetary masses $m_1=10^{-3}$~M$_\odot$ and $m_2=2\times 10^{-5}$~M$_\odot$, and
(ii) starting the two planets at a nominal closest approach, $\lambda_1=
\theta_{\rm min}$, $\lambda_2=0^\circ$ and $a_1=a_2=a_0=0.5$~au. Both were given
small initial eccentricity $e_1=e_2=0.001$, and the system was propagated for
1000~yr with the Burlish-Stoer integrator. We recorded series of planetary
transits and constructed a synthetic TTVs, similar to ones shown in Fig.~\ref{f3}.
The code also provided numerical mean values of the planetary orbital periods,
used to compute $P_0$, half-period $T$ of the TTV series, and the mean
value of the minimum planetary separation. This last parameter was obviously very
close to the given initial distance $\theta_{\rm min}$, but typically differed
from it by few tenths of a degree because of the effect of planetary eccentricities.
With those parameters determined for the direct numerical model, we have all
data needed to test the validity of Eq.~(\ref{link2}).
The results are shown by black circles in Fig.~\ref{f6}. The gray line is the
${\cal J}(\theta_{\rm min})$ integral from Eq.~(\ref{jint}), computed by a
Romberg's scheme with controlled accuracy. Note that this integration needs
a simple parameter transformation to remove the integrand singularity at
$\theta=\theta_{\rm min}$ limit. We note a very good correspondence of the 
numerical results with the expected trend from the analytic theory.
\begin{figure}[t]
 \epsscale{1.}
  \plotone{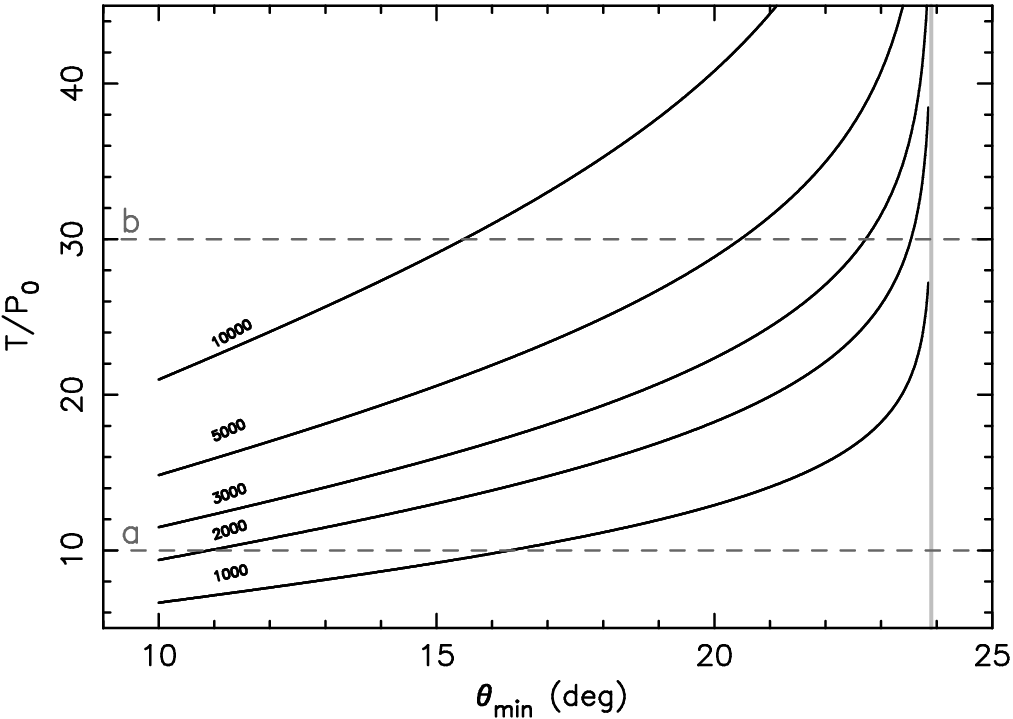}
 \caption{The observationally available ratio $T/P_0$, $T$ is half-period of
  the TTV series and $P_0$ orbital period, on the ordinate vs the minimum
  separation angle $\theta_{\rm min}$ of the planets during the coorbital
  cycle. The solid lines, evaluated using Eq.~(\ref{link2}), are given for
  five different values of the mass ratio $m_0/\sigma_+$ (labels). The gray
  line at $\theta_{\rm min}=23.9^\circ$ indicates the maximum theoretical
  value, while the true stability limit of $\theta_{\rm min}$ is lower as
  discussed in the text. The dashed gray lines a and b just set two examples
  of the $T/P_0$ values (see discussion in the text).}
 \label{f7}
\end{figure} 

Once we verified the validity of Eq.~(\ref{link2}), we can use it as shown
in Fig.~\ref{f7}. Here the abscissa is the minimum planet separation $\theta_{\rm min}$,
while the ordinate is now the ratio $T/P_0$ given for a set of different
$m_0/\sigma_+$ values (solid lines). The $T/P_0$ factor may be directly
constraint from the observations and Fig.~\ref{f7} hints that this information
may be immediately used to roughly delimit the $m_0/\sigma_+$ factor. This is because
$\theta_{\rm min}$ can span only limited range of value for the horseshoe
orbits: (i) $\theta_{\rm min}$ cannot approach too closely to the theoretical
limit $\simeq 23.9^\circ$ derived in Sec.~2.1, especially of $e_1$ and $e_2$
are non-zero, otherwise instability near the Lagrangian point L3 would onset,
and (ii) $\theta_{\rm min}$ cannot be too small, otherwise instability near the
Lagrangian points L1 and L2 would onset. While not performing a complete
study here, assume for sake of an example that $\theta_{\rm min}$ could be in
the interval $\simeq 10^\circ$ to $\simeq 22^\circ$. Then if $T/P_0=10$ is
obtained from the observations (as shown by the dashed gray line a in Fig.~\ref{f7}),
the $m_0/\sigma_+$ ratio cannot be much larger than $\sim 2000$. On the
other hand, in $T/P_0=30$ is obtained from the observations (as shown by the
dashed gray line b in Fig.~\ref{f7}), the $m_0/\sigma_+$ ratio cannot be much 
smaller than $\sim 3000$. Hence the observations may directly hint the nature
of planets in the coorbital motion.
\begin{figure}[t]
 \epsscale{1.}
  \plotone{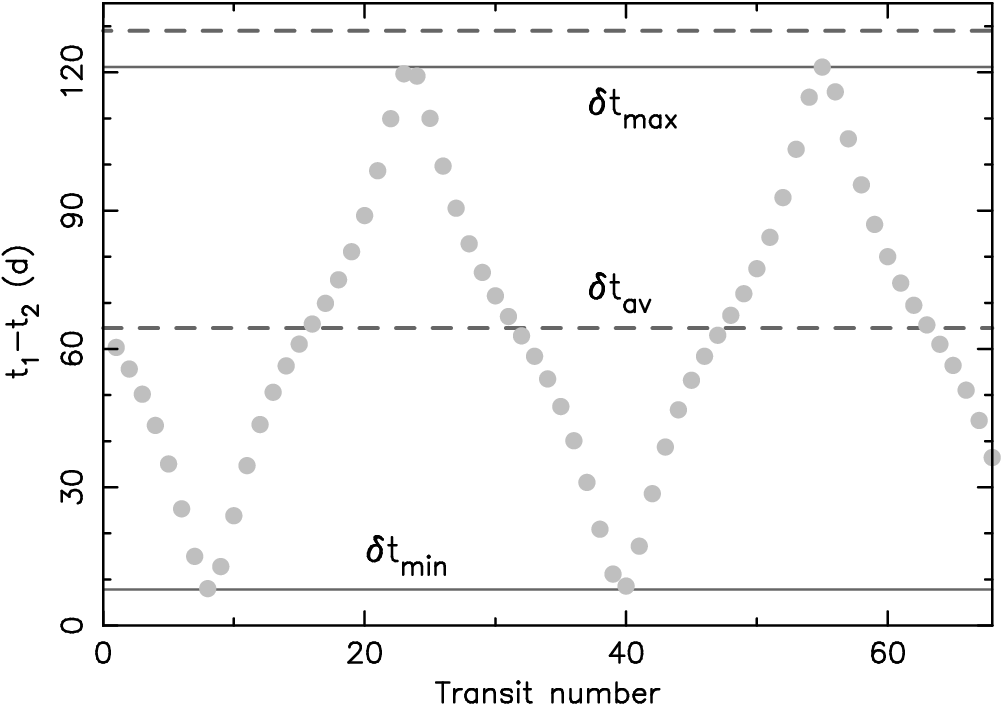}
 \caption{Difference between the transit time $t_1$ of the more massive
  planet and the immediately preceding transit time $t_2$ of the less massive
  planet in the simulation shown in Fig.~\ref{f3}. The abscissa is a transit
  number of the first planet. The average value of $t_2-t_1$, shown here as
  $\delta t_{\rm av}$, is half of the mean orbital period of the planets. The
  minimum and maximum values of $t_2-t_1$, shown here as
  $\delta t_{\rm min}$ and $\delta t_{\rm max}$, correspond to the extreme 
  configurations, when $\theta = \lambda_1-\lambda_2$ is either minimum or
  maximum.}
 \label{f8}
\end{figure} 

So far we discussed properties of TTVs obtained for the two planets. This is
because these series rely on transit observations of each of the planets
individually. While we have seen that a more complete information could be
obtained when we have TTVs for both planets, some constraints were available
even if transits of the larger planet are observed only. We now return
to the ideal case, when transits of both planets are observed and note
that even more complete information may be obtained by combining transit
series of both planets. Consider a series of transit instants $t_2$ of the 
second (less massive, say) planet and the consecutive transit instants $t_1$ 
of the first planet. We may then construct a series of their difference
$t_1-t_2$ as a function of the transit number. Using our example system
from above, this information is shown in Fig.~\ref{f8}. As expected, $t_1-t_2$
has the characteristic triangular, sawtooth shape which basically follows
from the time dependence of the planets' angular separation $\theta(t)$ and
spans values between zero and mean orbital period of the planets.
Consequently, the minimum value of $t_1-t_2$, which we denote $\delta t_{\rm min}$,
is directly related to the minimum angular separation $\theta_{\rm min}$ of the
planets. Similarly, the average value of $t_1-t_2$, say $\delta t_{\rm av}$, 
is one half the mean orbital period of the planets (and when $t_1-t_2\simeq 
\delta t_{\rm av}$, the planets are at opposition). As a result
\begin{equation}
 \theta_{\rm min}/\pi = \delta t_{\rm min}/\delta t_{\rm av}\; .
\end{equation}
Since $\delta t_{\rm min}$ and $\delta t_{\rm av}$
are in principle discerned from observations, $\theta_{\rm min}$ can be
fairly well constrained as well. In the same way, the maximum value of $t_1-t_2$, say $\delta 
t_{\rm max}$, provides 
\begin{equation}
 \theta_{\rm min}/\pi =2-\delta t_{\rm max}/\delta t_{\rm av}\; .
\end{equation}
As an example, $\theta_{\rm min}$ estimated from the series in Fig.~\ref{f8} is
$\simeq 21.9^\circ$, which is very close to the numerically obtained value of
$\simeq 21.4^\circ$. With $\theta_{\rm min}$ constrained, we note that the
TTVs analysis using Eq.~(\ref{link2}) provides an independent, correlated constraint
of $\theta_{\rm min}$ and the normalized sum of planetary masses $\sigma_+/m_0=(m_1+
m_2)/m_0$. Henceforth, $\sigma_+/m_0$ can be directly obtained. If combined with
the information about their ratio $m_1/m_2$ discussed above, we finally note that
individual planetary masses $m_1$ and $m_2$ (given in $m_0$ units) can be
determined from the observations.

\section{Conclusions}\label{concl}
While still awaiting for the first confirmed exoplanetary coorbital configuration,
we derived here simple parametric relations that could be revealed from the
TTV series of a such a system. From all possible coorbital architectures
we chose here the horseshoe case that provides TTVs having the most
singular nature. This is because at the zero order one may consider this
situation as two non-interacting planets that periodically switch their
orbits around some mean distance $a_0$ from the star. Instead of sinusoidal in
nature, the TTVs thus resemble a triangular-shaped series with the half-period
$T$ of the coorbital motion. 

In an ideal case, where TTVs of both planets are observed, we find that the
characteristics of a complete-enough dataset of planetary transits may directly 
provide information about their masses. This is because the ratio of the TTV amplitudes 
constrains directly the ratio of the planetary masses. Additionally, time separation
between the transits of the two planets allows to constrain their minimum angular
separation $\theta_{\rm min}$ as seen from the star. This information, if combined 
with Eq.~(\ref{link2}),
then provides a constraint on the the sum $\sigma_+=m_1+m_2$ of the planet's masses
in units of the stellar mass $m_0$.

Even if TTVs of only larger coorbiting
planet are observed, say, one may use Eq.~(\ref{link2}) to relate the total mass of
planets, $\sigma_+$, to their minimum angular separation $\theta_{\rm min}$.
This only requires the data constrain $T$ and the mean orbital period $P_0$,
or rather their ratio $T/P_0$. Since the available range of $\theta_{\rm min}$
value is limited for stable orbital configurations, the value $T/P_0$ itself
roughly sets a possible range of planetary mass, allowing us to distinguish cases
with Jupiter-mass planets as opposed to the super-Earth-mass planets 
participating in the coorbital motion.

\acknowledgements
 The work of DV was supported by Czech Grant Agency (grant P209-13-01308S). We
 thank an anonymous referee for useful suggestions on the submitted manuscript.

\end{document}